# Controlled Polar Asymmetry of Few-Cycle and Intense Mid-Infrared Pulses

Christian Schmidt,[1] Johannes Bühler,[1] Bernhard Mayer,[1] Alexej Pashkin,[1] Alfred Leitenstorfer,[1] and Denis V. Seletskiy[1,*]

[1]Department of Physics and Center of Applied Photonics, University of Konstanz,
Universitätsstraße 10, 78464 Konstanz, Germany
*Corresponding author: denis.seletskiy@uni-konstanz.de

**We demonstrate synthesis of super-octave-spanning and phase-locked transients in the multi-terahertz frequency range with amplitudes exceeding 13 MV/cm. Sub-cycle polar asymmetry of the electric field is adjusted by changing the relative phase between superposed fundamental and second harmonic components. The resultant broken symmetry of the field profile is directly resolved via ultrabroadband electro-optic sampling. Access to such waveforms provides a direct route for control of low-energy degrees of freedom in condensed matter as well as non-perturbative light-matter interactions.**

With the availability of intense and stable femtosecond sources, precise coherent control of light-matter interactions has been shifting into the focus of ultrafast science [1,2]. This step necessitates the ability to generate tailored phase-stable optical waveforms with sub-cycle precision [3]. Some prominent examples constitute synthesis of single-cycle pulses [4, 5] or novel fields with symmetry-broken polarity on sub-cycle and envelope timescales. The most direct approach to accomplish the latter is to superimpose a fundamental field with its phase-locked second harmonic, while exercising control over the relative phase between these components. It has already been demonstrated that synthetic two-color pulses can be used to drastically enhance generation of THz [6-9] radiation in gas plasmas via off-resonant interaction. In addition, quantum interference of one- and two-photon resonant absorption pathways enables all-optical injection of ballistic charge and spin currents in bulk [10-12] and low-dimensional [13-15] semiconductors. Together with coherent control of molecular anisotropies [16], all those examples are governed by the resulting polar asymmetry in the envelope of the hybrid pulse [13,17-20]. On the other hand, non-perturbative light-matter interactions in attosecond [21-23] or solid-state [24] physics require asymmetry on sub-cycle timescales which can also be accomplished via harmonic synthesis [9,25,26].

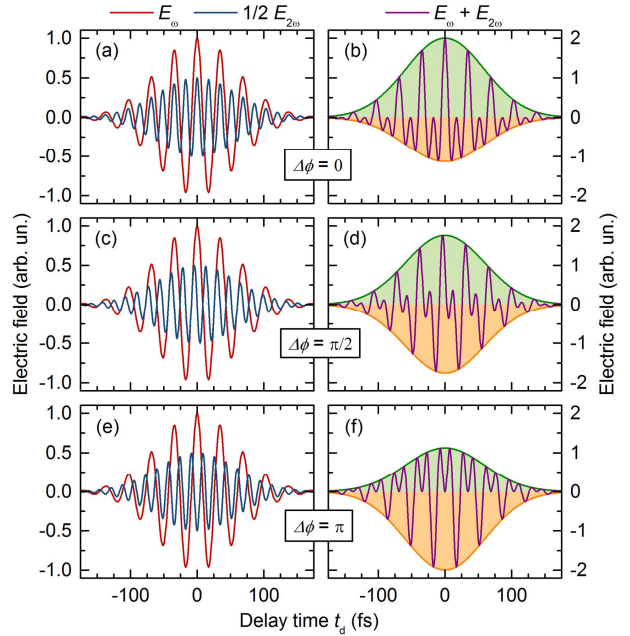

Fig. 1. (Color online) Calculated synthetic waveforms (right column) consisting of a fundamental (30 THz, red) and second harmonic (60 THz, blue and rescaled for clarity) constituent fields (left column) for three different relative phases $\Delta\phi$ of 0 (a,b), $\pi/2$ (c,d) and $\pi$ (e,f). Shading highlights direction of the polar asymmetry of the synthetic transient, controlled by the $\Delta\phi$ parameter.

In this Letter we demonstrate generation and field-resolved detection of synthetic mid-infrared (MIR) transients with strongly asymmetric polarity and peak amplitudes up to 13.9 MV/cm. Coherent superposition of fundamental and second-harmonic (SH) pulses is exploited. Adjustment of the relative phase between these two components allows for direct control of the degree of asymmetry on both the sub-cycle and envelope timescales.

From symmetry considerations, synthesis of a waveform with unbalanced polarity requires a superposition of a fundamental field and an even harmonic, in the simplest case the SH frequency. The total field can be represented as a sum of fundamental (amplitude $E_1$ at the carrier angular frequency $\omega$) and SH ($E_2$) components

$$E_{\text{tot}} = \Re\left\{ E_1 e^{-i\omega t} + \gamma E_1 e^{-i2\omega t} e^{i\Delta\phi} \right\}, \qquad (1)$$

where $\gamma = E_2/E_1$ and $\Delta\phi$ define the amplitude ratio and the relative phase difference of the two fields, respectively. To illustrate this point, Figure 1 depicts a calculated $E_{\text{tot}}$ and its constituents for three representative cases of $\Delta\phi = 0$, $\pi/2$ and $\pi$. For $\Delta\phi = 0$, the synthesized field alternates between constructive and destructive interference with each successive period of the fundamental wave. While the time-average of the field is $<E_{\text{tot}}> = 0$, the resultant waveform is nonetheless endowed with broken polar symmetry (Fig. 1 b) involving, for example, a finite value of $<E^3_{\text{tot}}>$ [27]. For the $\Delta\phi = \pi/2$ case, polar symmetry is restored as both constituents are now out of phase (Fig. 1 d). Finally, a relative phase $\Delta\phi = \pi$ (Fig. 1 f) results in a waveform with a reversed polarity in comparison to $\Delta\phi = 0$. Thus, sub-cycle asymmetry can be directly manipulated by changing the relative phase between the fundamental and SH fields.

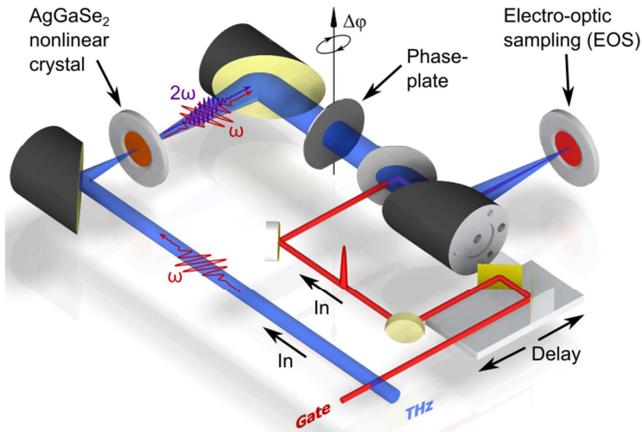

Fig. 2. Experimental setup for synthetic MIR pulse generation and detection. Phase-stable multi-THz pulses of in-plane polarization at the fundamental angular frequency $\omega$ are focused into the SH crystal (AgGaSe$_2$, type-I phase matching, see text for details), generating a $2\omega$ transient. A GaSe crystal, placed in the collimated part of the multi-THz optical path, serves as a phase plate to control the relative delay between the two field constituents. An 8 fs gate pulse and a 10-µm-thick <110>-oriented ZnTe crystal enable ultrabroadband detection via electro-optic sampling. The temporal offset of the gate pulse with respect to the multi-THz waveform is controlled with a variable optical delay.

The schematic of the setup is depicted in Figure 2. A frequency-doubled amplified pulse train from one of the two branches of our femtosecond Er:fiber master oscillator is used for seeding a Ti:sapphire regenerative amplifier [28]. Broadly-tunable few-µJ pulses in the multi-THz frequency range (blue beam path in Figure 2) are derived by difference frequency mixing of signal beams from mutually-synchronized optical parametric amplifiers (OPAs) pumped at 800 nm [28,29]. A center frequency of approximately 30 THz was chosen in our experiments to ensure the possibility of broadband electro-optic sampling (EOS) of both fundamental and SH components. Frequency doubling is achieved in a 350-µm-thick silver gallium selenide (AgGaSe$_2$) nonlinear crystal ($\theta = 56°$, $\varphi = 45°$) [30], placed at an intermediate focus (full width at half maximum of intensity: 60 µm) of the THz pulse train. The emerging phase-locked electric fields at carrier angular frequencies of $\omega$ and $2\omega$ are re-collimated and transmitted through a thin dispersive plate, serving as a phase retarder for adjustment of $\Delta\phi$. In our experiment, this control element is implemented by a gallium selenide (GaSe) crystal with a thickness of 580 µm. Small azimuthal rotation of the plate introduces a relative path-length difference between the two components of the hybrid field, inducing a commensurate phase slip between them [18]. The resulting synthetic waveform is monitored by an ultrabroadband electro-optic sampling (EOS) technique, employing a <110>-oriented zinc telluride (ZnTe) detector crystal of a thickness of 10 µm [31,32]. An 8-fs near-infrared sampling pulse (red beam path in Figure 2) is derived from a parallel Er:fiber amplifier branch after the master oscillator [28]. We rotate the ZnTe sensor to ensure the same ratio of the detected orthogonally-polarized $\omega$ and $2\omega$ field amplitudes as present after the SH crystal, in analogy to the effect of a linear polarizer with variable polar angle.

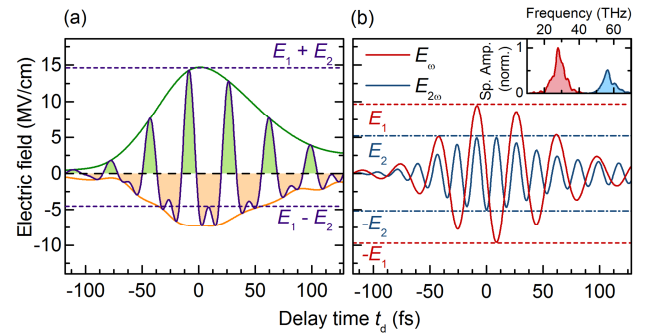

Fig. 3. (a) Analysis of a synthesized two-color field transient versus delay time $t_d$. Polar asymmetry of the electric field is highlighted with green and orange shading for positive and negative amplitudes, respectively. The peak magnitude of the resultant field of 13.9 MV/cm has been derived from the measured pulse parameters of the $\omega$ beam and the detector geometry. (b) Fundamental (centered at 28 THz) and second harmonic (56 THz) wave packets, as obtained from the inverse Fourier transform of the spectral components (inset). Color-coded dashed horizontal lines mark the $E_1$ and $E_2$ maxima in the peak amplitude of the $E_\omega$ and $E_{2\omega}$ waveforms, respectively. The calculated limits of full constructive ($E_\omega + E_{2\omega}$) and destructive ($E_\omega - E_{2\omega}$) interference are depicted as dashed lines in panel (a). Strong agreement of the measured envelope to these limits suggests nearly ideal temporal matching between the two fields.

The measured total electric field of the two-color transient is depicted in Figure 3 a. The envelope of the synthetic waveform exhibits a strong unipolar component arising from constructive interference between positive excursions of the field cycles of the two harmonics. In particular, the positive polarity of the envelope (green solid line) is more than two times higher as compared to the one in the negative direction (orange solid line). Green and orange highlights underscore the pronounced asymmetry between positive and negative polarities on sub-cycle timescales, respectively. Whereas the integral over positive and negative field values of one cycle are approximately the same, the extrema exhibit an amplitude polarity of more than a factor of two. Direct Fourier analysis identifies the spectral content of the waveform (inset in Fig. 3 b) to consist of broadband fundamental (28 THz) and second harmonic (centered at 56 THz) components. The two constituents are isolated in the spectral domain and transformed back into the time domain via a fast inverse Fourier transform (Fig. 3 b). We determine the amplitude of the SH (blue solid curve) field to be approximately half as compared to the fundamental (red solid curve). This finding corresponds to an energy efficiency in SH generation of 25%, as confirmed by an ultrabroadband study of the spectral content based on linear autocorrelation (not shown). Thus, optimal phase-matching conditions in AgGaSe$_2$ result in a short and intense SH pulse supporting strong polar asymmetry. It is interesting to note that the extracted peak amplitudes of $E_\omega$ and $E_{2\omega}$ can be used to form the expectation bounds for perfect constructive ($E_\omega + E_{2\omega}$) and destructive ($E_\omega - E_{2\omega}$) interference in the total measured field in Fig. 3 a. Excellent agreement between the measured extrema of the envelope of the two-color pulse and the theoretical bounds highlights the level of precision in temporal matching between the mutually coherent field constituents. In the experimental trace, the in-phase oscillation of SH maxima and the extrema of the fundamental field correspond to a relative phase between the two inputs of $\Delta\phi \approx 0$. Theoretical minima ($E_1 - E_2$) and maxima ($E_1 + E_2$) of the synthetic field (horizontal indicators in Fig. 3 a) can be used to define the degree of asymmetry $\eta = 1 - (E_1 - E_2)/(E_1 + E_2) = 2\gamma/(1+\gamma)$, such that $0 \leq \eta \leq 1$. Accordingly, the degree of asymmetry of the two-color waveform is determined from the experiment to be $\eta = 0.68$, very close to the maximally expected value of $\eta = 0.69$ for the measured ratio of the field amplitudes of $\gamma = 0.52$ (Fig. 3 a).

Next, we demonstrate direct manipulation of the sub-cycle polar asymmetry by variation of the relative phase $\Delta\phi$ between the two components with the GaSe phase plate. The generated fields are monitored in the time domain and with sub-cycle resolution. Figure 4 displays three traces of the total synthesized wave packets for different settings of the phase plate, corresponding to three different relative delays. Measured and calculated data for $\Delta\phi = 0, \pi/2$ and $\pi$ are depicted. Here we focus on the sub-cycle structure of the synthesized field by zooming onto the central part of the multi-THz temporal profile. The change of polar asymmetry outlined in Figure 1 is now implemented experimentally. One directly observes how the relative phase influences the field profile of the measured pulse: A strong unipolar peak in field is produced for $\Delta\phi = 0$, followed by symmetric field excursions with a cubic-like shape at zero-crossings for $\Delta\phi = \pi/2$, and finally a unipolar peak but with reversed polarity for $\Delta\phi = \pi$. Hence, the sub-cycle shape of the two-color waveform can be completely controlled by adjusting the relative phase $\Delta\phi$ between the two inputs. With peak fields in excess of 13 MV/cm and synthesized polar asymmetry, such pulses are ideal candidates for optical control over charge and spin currents in material systems with simultaneously allowed one- and two-photon absorption [33-35], as well as non-perturbative light-matter interactions [36].

In summary, we demonstrated synthesis and field-resolved detection of intense MIR transients with controlled polar asymmetry. Collinear and efficient SH generation in AgGaSe$_2$ was exploited to generate waveforms spanning from 20 to 70 THz. Extended arrangements can offer additional flexibility such as independent amplitude, polarization and dispersion control of the constituents. These waveforms can be readily applied for quantum control of low-energy elementary excitations in condensed matter with sub-cycle time resolution.

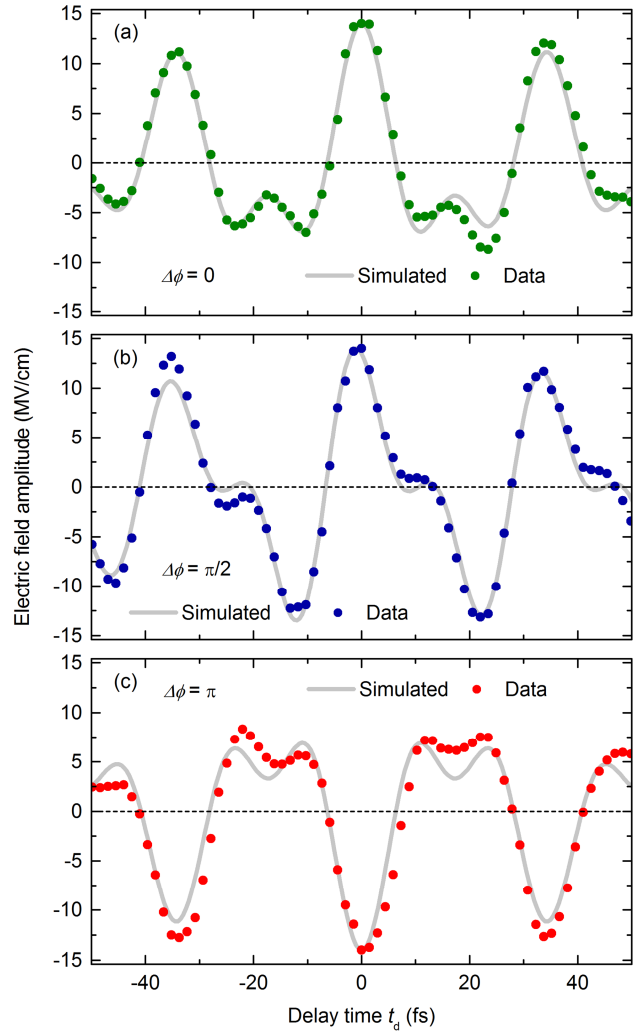

Fig. 4. (a) Sub-cycle structure of the electric field of synthetic two-color mid-IR waveform for three different phase configurations. The electric field amplitude is displayed over delay time $t_d$ for EO detected data points (colored points) and calculation following Eq. 1 (grey solid line). In (a) the phase difference between fundamental and SH field is set to $\Delta\phi = 0$. Whereas for (b) and (c) the relative phase is changed to $\Delta\phi = \pi/2$ and $\Delta\phi = \pi$, respectively.